# Interpretable and Actionable Vehicular Greenhouse Gas Emission Prediction at Road link-level

**S.Roderick Zhang**

Laboratory of Innovations in Transportation (LiTrans)
Ryerson University, Toronto, Canada
s11zhang@ryerson.ca

**Bilal Farooq**

Laboratory of Innovations in Transportation (LiTrans)
Ryerson University, Toronto, Canada
bilal.farooq@ryerson.ca

**Abstract**

To help systematically lower anthropogenic Greenhouse gas (GHG) emissions, accurate and precise GHG emission prediction models have become a key focus of the climate research. The appeal is that the predictive models will inform policymakers, and hopefully, in turn, they will bring about systematic changes. Since the transportation sector is constantly among the top GHG emission contributors, especially in populated urban areas, substantial effort has been going into building more accurate and informative GHG prediction models to help create more sustainable urban environments. In this work, we seek to establish a predictive framework of GHG emissions at the urban road segment or link level of transportation networks. The key theme of the framework centers around model interpretability and actionability for high-level decision-makers using econometric Discrete Choice Modelling (DCM). We illustrate that DCM is capable of predicting link-level GHG emission levels on urban road networks in a parsimonious and effective manner. Our results show up to 85.4% prediction accuracy in the DCM models' performances. We also argue that since the goal of most GHG emission prediction models focuses on involving high-level decision-makers to make changes and curb emissions, the DCM-based GHG emission prediction framework is the most suitable framework.

*Keywords:*  Greenhouse Gas Emission Prediction, Discrete Choice Model, Decision-Making, Urban Road Network,

# 1. Introduction

Transportation sector has been working on different ways to lower Greenhouse Gas (GHG) emissions from road networks to mitigate anthropogenic climate change and ensure sustainable development. One commonly discussed solution is to replace fossil fuel-dependent vehicles with alternative-fuel-based green vehicles. While those alternative-fuel vehicles with low vehicular GHG emissions are gaining popularity amongst vehicle owners, a large majority of on-road vehicles are still running on fossil fuels. In North America, private vehicles remain the predominant mode of transportation. Within this market, there is a significant gap in terms of purchase and operation cost between alternative-fuel-based vehicles and traditional fuel-dependent vehicles for most owners. Thus, until the manufactures and alternative-fuel providers can cut down the cost to a level that is competitive with traditional fuel-driven vehicles' associated cost and ensure effective infrastructure supports are in place (e.g fast charging stations for electric vehicles), the commercial green vehicle[1] has a long way to go to populate all the road network in North America. In November 2020, the Standing Committee on Environment and Sustainable Development of Canada publicly recognized that the nation has fallen off track to reach the target of having 100% of all light-duty cars be electric by the year 2025 on Environment and Development (2021). Consequently, Canada's ambitious goal of reaching 100% electric light-duty cars by 2040 becomes less likely to be achievable in time. As a result, alternative means to lower GHG emissions are required from the transportation sector during this time of transition. System-based methods such as alleviating traffic congestion to minimize idling vehicles and maximizing the utility of passenger vehicles by reducing the number of empty trips are both effective strategies to bring down vehicular GHG emissions. On the user-based side, emission-conscious trip planning technologies such as eco-routing solutions, e.g. Djavadian et al. (2020b), is another potential way to lower the GHG emission produced by fossil-fuel-driven vehicles before the age of zero-emission vehicles arrive. Nevertheless, regardless of which type of approach is involved, to address any issues related to the transportation sector we must recognize that transportation is, first and foremost, a public service. To be able to truly achieve any substantial success with curbing GHG in transportation, we must involve government decision-makers and address the problem at a societal magnitude from a high level. In this work, we seek to establish a framework of an interpretable and actionable prediction model of GHG emissions at the link level of road networks. The model would help decision-makers address the key factors contributing to high GHG emissions from a high level with ease and confidence. The key contributions of this paper are:

- First attempt on using Discrete Choice Models to directly make GHG emission level predictions at link (road segment) level in urban road networks.

- Established an analysis framework to produce interpretable predictions that induces direct actions from decision-makers

---

[1]A green vehicle is a clean, eco-friendly or environmentally friendly road motor vehicle that produces less harmful impacts to the environment than conventional internal combustion engine vehicles running on gasoline or diesel, or one that uses certain alternative fuels.

The rest of the paper is divided as follows: In the background section, we review various GHG emission prediction methods with a focus on the ones popular in the transportation sector. We then introduce the main analysis method used in this work. In the Method and Data section, we go into the specifics of the analysis with details on the models that we use. We also explain the data used in the analysis here. In the following result and discussion section, we present the model results and discuss their significance. Finally, we conclude with a summary of the highlight and some potential outlooks for the future.

## 2. Background

In this section, we examine some of the well-established and relevant Greenhouse Gas prediction methods for our purpose of identifying the most suitable and effective prediction model for higher-level decision-makers. We start by reviewing the most popular GHG prediction method, namely neural network-driven models, from a descending scale of large global-wide study to local urban application. Next, we look into a variety of domain specific methods that have been widely adopted in the transportation sector. Finally, we explore the potential of using Discrete Choice Modelling, which is more often used in socio-economic context of informing policy in applications to climate change studies, as a direct prediction tool for Greenhouse Gas emissions.

*2.1. Greenhouse Gas Emission Prediction using Neural Networks*

Due to the increased awareness of climate change and its negative effects, extensive research has been going on the prediction of GHG emissions from various sources. Emission prediction models can be an effective tool for understanding the process and major influencers and developing proactive measures. Of all the prediction methods, Neural Network (NN) is the most popular and one of the most effective methods used Abdullah and Pauzi (2015). NN gained popularity for its robustness, variety of algorithms, minimum understanding of the data required from the user, and its simplicity to implement for less experienced programmers.

On a global level, Behrang et al. (2011) used socio-economic indicators as the base and conducted a forecast of world CO2 emissions. The specific indicator used is global population, GDP, oil and natural gas trade movement from a 27-year (1980-2006) interval. The model framework is based on an integrated multi-layer perceptron neural network and Bees Algorithm (BA). The study used BA to come up with the demand equations (linear and polynomial) for fossil fuel and primary energy using the aforementioned indicators. A comparative analysis was performed for the two fitting methods to obtain the future socio-economic indicators. In the first case, the study fitted multiple polynomial trend lines to the observed indicators and the best-fitted one with the highest R-square was kept forecasting the future values of the indicators. In the second approach, multiple NNs were trained on the observed indicators, and the best performing NN was used to predict future indicator values. As a result, the multi-layer NN outperformed the multiple polynomial trend line fitting, and the predicted future indicators from the best performing NN were fed into the demand equations. Based on the future fossil fuel and primary energy demand, future emissions of $CO_2$ were then projected.

At the scale of a country, Radojević et al. (2013) used the NN technique for predicting the environmental indicators of sustainable development in Serbia. The method was introduced mainly as a solution to enable the team to forecast with an incomplete dataset. Similar to Behrang et al., the study also used socio-economic indicator data such as gross domestic product (GDP), gross energy consumption, and energy intensity data from 1999 to 2007 as the input parameters. Various development scenarios were simulated to predict the GHG emission up to 5 years into the future. The relative error of the predicted GHG was less than 10%. The obvious takeaway from this study is that NNs are a suitable method for predicting GHG emissions. More importantly, the study showed that NN models can be used to simulate various societal development scenarios with different national-level socio-economic indicators as predictors, so that governments and industry's action impact can be evaluated and, in turn, help with decision-making on sustainable development.

Aiming to apply the advanced Neural Network-based GHG prediction to the specific sector of transportation, Alhindawi et al. (2019) applied an Adaptive Neuro-fuzzy inference system (ANFIS), which incorporates both neural networks and fuzzy logic principles to deal with non-linear data, to model and forecast GHG emissions from the North American road transportation. The input data used were GHG emission over 20 years, the ratio of vehicle kilometers by mode, and the number of vehicles. The analysis shows that all different modes of transport contributed equally to the model prediction, proving the reliability of the ANFIS model in optimizing the given data to predict GHG emissions from the road transportation sector.

Another Neural Network-based model that is capable of handling nonlinear data is the Long-Short term memory model, which is a type of Recurrent Neural Network. Focusing on the road network level of a city, Alfaseeh et al. (2020) developed an auxiliary LSTM that aims to better predict GHG emission rate on links (roads) by overcoming the limitations of low spatial and temporal resolution in the data, so that in turn, the LSTM based GHG emission prediction can be used to optimize eco-routing accuracies. A set of sufficiently large predictor data were generated from MOVES [2] and microsimulation of traffic in the downtown Toronto. The study cross-compared amongst three different models: autoregressive integrated moving average (ARIMA), Clustering (K-means), and auxiliary Long-Short term memory (LSTM), and found that when combined with Bayesian optimization/updating, the LSTM (two-layers) outperforms the other two methods in terms of predicting the link-level GHG emission rate. It also performed correlation analysis to determine the most important independent variables (indicators/predictors) to use, namely speed, density, in-link flow, and GHG emission rate.

---

[2] Motor Vehicle Emission Simulator

*2.2. Transportation Emission Estimation based on Causal Relations*

There is a large collection of studies that estimate the emission from the transportation sector where the authors focused on predicting variables that have known causal relations with emissions. Once the primary endogenous variable's value has been determined, the authors then feed the predicted variable values as input to established frameworks to estimate the emission. These approaches are only justifiable if a known causality has been established between the variables selected and emission.

For instance, in the previous section, we saw that Behrang et al. (2011) and Radojevic´ et al. (2013) both used socio-demographic indicator variables as the base of GHG prediction. Their approaches are justifiable because other case studies have found supporting evidence that Socio-Economic factors are indeed associated closely with greenhouse gas emission. Pao et al. (2014) conducted a study to model emissions and energy consumption with economic growth in Brazil. The study found that there is a bidirectional strong causality running between income, energy consumption, and Carbon Dioxide emission. Ameyaw and Yao (2018) also conducted a causality analysis in five west African countries, revealing that there exists a unidirectional causality going from GDP to Carbon Dioxide. The study continued to implement a bidirectional LSTM to predict the Carbon dioxide emission based on GDP and achieved a prediction accuracy of above 90%, further justifying the use of socioeconomic indicators as input variables.

In transportation, extensive studies have gone into vehicle fuel consumption predictions and identifying key factors that contribute to fuel efficiency. The interest arises from the concept that fuel amount can be converted into emission amount, for carbon amounts are measurable in both polluted emissions and in fuel. Since fuel consumption has a known causal relation to vehicle emissions Churkina (2008); Stanek and Breiland (2013); Ameyaw et al. (2019), many fuels efficiency-based studies were able to then deduce the number of potential vehicle emissions based on the exhausted fuel and similarly estimate the contributing factor to vehicle emissions by inference.

Zhao et al. (2011) used cluster computing on real-time data collected by the Intelligent Traffic System to generate annual data of fuel combustion and then in turn predicted the emission data using the MapReduce framework. Yao et al. (2020) built a fuel consumption model using a backpropagation neural network, support vector regression, and random forests. The dataset was based on driving behaviour collected by a mobile phone as well as actual fuel data collected by an onboard diagnostic system. The findings indicate random forest to be the most robust and accurate method. Delussu et al. (2021) deployed heuristic and exhaustive algorithms to generate Bayesian networks among sensor monitored traffic and environment data to predict the fuel consumption dependent on the context of traffic conditions. In another inferred case of predicting consumption, Rito et al. (2021) extracted vehicle count data from google map and deduced the energy consumption data. Traffic variables dataset was enriched with the consumption data and multiplied assumed emission factors to estimate the emissions.

Another popular area of study that has a known causal correlation with transportation-induced emission is the study of vehicle operation. Gallus et al. (2017) analyzed the impact of driving style and road grade on emissions of passenger vehicles. The authors found that cumulated altitude gain and road grade was directly correlated to emissions of Carbon Dioxide and Nitrogen Oxides, where a step increase from 0 to 5% in road grade leads to a 65-81% increase in Carbon Dioxide and an increase of 85-110% in Nitrogen Oxides. In terms of estimating the emission, popular vehicular emission models such as MOVES, CMEM[3], and IVE[4] depend on vehicular operating conditions to make the prediction. Hence the more information one has on the different operating conditions that the vehicles spend within, the more accurately one can predict emission rate values. Fan and Daniel (2016) used VISSIM[5] simulation software to generate second-by-second running activities data of vehicles in five different operation modes at a signalized intersection. The operation condition data were then used as input for MOVES to estimate the emissions. The findings of this work established the significance of operation modes on Carbon Monoxide emission and established a basis for understanding the impact that operation mode has on pollutant emissions. By aggregating Emissions Specific Characteristics (ESC) such as road geometry, road environment, traffic conditions, and driving behavior, Nesamani et al. (2017) was able to more accurately predict the fraction of time vehicles spend in various operating conditions. The study also analyzed the impacts that different ESCs have on vehicle operation. The main significance of this study is a new vehicle operation estimate model which provides traffic agencies and practitioners with a way to improve emission estimates.

---

[3]Comprehensive Modal Emission Model
[4]International Vehicle Emissions model
[5]Verkehr In Städten - SIMulations Model (German for "Traffic in cities - simulation model")

All the methods and approaches discussed so far have proven to generate decent emission predictions, with most of the studies mentioning that better emission estimation leads to better-informed decision (often policy) makers. These decision-makers can, in turn, curb down the emissions through appropriate series of actions. However, if the eventual goal of predicting emission to high accuracy with robustness is to inform decision-makers, then the researchers would have to consider the aspect of interpretability and actionability of their models. All the aforementioned predictions focused on predicting continuous emission data, and while some of them were able to achieve very high accuracies using very complex structures, the continuous value results are not immediately interpretable to the decision-makers to enable their immediate actions. In the following section, we will introduce a versatile model, tailored for decision-makers, that is not only attractive for its interpretable and actionable analysis but also capable of predicting emission in a parsimonious fashion.

*2.3. Discrete Choice Analysis and Environmental Decision Making*

Discrete Choice Models (DCM), as the name suggests, are built to be applied to and analyze a broad range of aspects that concerns the probability of choosing certain specific alternatives that are of discrete data type. Fundamentally, DCMs predict the likelihood of specific alternatives, given meaningful explanatory variables/attributes. Additionally, Discrete Choice Analysis (DCA, used interchangeably with DCM) can also be a convenient tool to perform sensitivity analysis. The usefulness of sensitivity analysis is that it allows analyzation of key impact factors contributing to the alternatives. Furthermore, direct computation of the change in the endogenous variable based on the changes in its predictors' values can also be conducted conveniently due to DCM's linear nature. All these functional advantages make DCM a very informative and user-friendly tool for decision-makers who are looking for actionable solutions. We now present some of the relevant examples of DCM applications.

In the theme of greener maritime transportation, Franchi and Vanelslander (2021) recently conducted a study pursuing to solve environmental problems in the world of maritime transport by establishing greener strategies in ports. The study conducted a revealed preference survey with 14 container shipping companies to collect the data and then adopted the DCA approach to analyze the most important criteria of influence to make a port green and to highlight the key ones for container shipping companies when deciding their approach to ports. Using a multinomial logit (MNL) model, the study identified that the most impactful regulation attributes that can help make a port greener are air pollution level, port capacity, and productivity, as well as total cost and charges.

Traditionally, DCA is often applied to human preference and behavior data like the reveal preference survey in Franchi and Vanelslander (2021). However, using human-centric data does not always have to be the case to apply DCA. In a climate-change-related land use analysis, Cho and McCarl (2021) used historical satellite data of U.S. farmlands to simulate the projected land use allocations in 2030, 2050, and 2070 with CMIP5. The study deployed a two-step linearized multinomial logit model to examine the impacts of spatial factors has on land use, as well as the impacts of climate factors. Making the land the decision-maker in the first step and climate the decision-maker in the second step. The analysis found that spatial factors such as the use of the nearby land have a notable impact on the land use of the target area. It also found that for climate factors there are thresholds above which temperature or precipitation causes cropland shares to diminish and pasture/grass shares to increase, showing an opposite yet significant effect these climate attributes have on crop versus pasture/grass shares. The study concluded with implications to help policymakers tackle climate-driven land-use changes and farmers adapt to climate change.

In this work, we will make the links in the road network the "choice-maker" [6] (similar to the non-human centric decision-maker approach of Cho and McCarl (2021)) to conduct a DCA analysis and identify the most impactful link-related factors contributing to high GHG emission rate in the downtown road network of Toronto to help inform actions. We will also discretize the continuous GHG emission rate data into low-medium-high categories and make categorical likelihood predictions of GHG emission level with DCM. By discretizing the GHG data, we simplify the complexity of predicting continuous emission value without losing any significance of the information contained in the data. Most of the advanced emission prediction models, like some of the ones mentioned earlier, adopt more and more sophisticated structures to improve their prediction precision to the decimals, while compromising interpretability. However, for high-level decision-makers who look to curb down emissions, the precision of the exact emission value predicted is not going to change their perception of what actions to take. As far as decision-makers are concerned, they only need to know if the GHG emission is at a high level, which urgently needs to be addressed, or if the GHG emission is at a tolerable medium level which can be addressed now if necessary, or if the GHG emission is low enough that no changes are required from the decision-makers. Addtionally, they would also want to know supporting information such as what are the key traffic, built-space, and geometric variables that contribute and affect the GHG levels the most. Therefore, it is unnecessary to use complex and expensive prediction methods to pursue the decimals if the main purpose is to inform decision-makers when a parsimonious model could suffice. To the best of our knowledge, this is the first attempt to adopt DCM to predict link-level GHG emission on a road network. The details of the implementations to our approach will be discussed in the methodology section.

---

[6]This is not to be confused with decision-maker. Choice-maker is defined from the model's perspective, decision maker is from the user's perspective, who would be using the DCA as a tool.

## 3. Methodology

The methodology section will mainly cover the mechanism and specification related to Discrete Choice Model, along with an introduction to the data we are using, as well as the processing that went into preparing the data. In section 3.1 we explain the mechanism of DCM. Examples of different DCMs are also given with reference to some prior studies as examples to illustrate different applications. In section 3.2 we go into the details of the data we are using and the processing methods we performed on the data. Finally, in section 3.3 we talk about the specifics of the set-up for our DCM models.

### 3.1. Discrete Choice Modelling

In this study, we seek to establish that DCM is not only capable of predicting link emissions levels in a parsimonious, interpretive and effective manner, but it is also an ideal tool for decision-makers to cut down emissions in road networks on a larger societal scale. To show the capability of DCM for our tasks in a clear manner, the main discussions are centered around the MNL model since it is the most iconic and fundamental model of all the DCMs. The standard choice probability equation is shown in Equation 1. $Y_i$ represents a discrete choice among $J$ alternatives, and the terms $\beta_j X_i$ together is the expected utilities. Here, $\beta_j$ is the regression coefficients, and $X_i$ is the attribute of individual. In the standard MNL model, the same attribute $X_i$ is used to model the utilities of all $J$ alternatives. We should note that there are other niche specific DCM models such as the ordered logit model, which we also implemented to obtain supporting evidence for the DCM models' results, that are useful in capturing ordinal ranking attributes in the response variable to help improve the performance. To shed some light on the capability of some of those more niche-specific models, we will mention a few relevant works that share some similarities with our setup and utilized models that we believe could be useful to us in future works.

$$Pr(Y_i = J) = \frac{1}{1 + \sum_{j=1}^{J-1} e^{\beta_j X_i}} \quad (1)$$

The ordered logit model (standard equation shown in Equation 2) is an ordinal regression model, which is a type of regression model used for ordinal dependent variables (McCullagh, 1980). Of all the types of the ordered logit model, the most common model is Proportional Odds Model (PO). It compares the probability of response less than or equal to a given category to the probability of a response greater than this category. Similar to MNL models, for a total category of C, the ordered logit model is also composed of C-1 equations. When the total category C is equal to 2, it behaves just like a classic (binary) logistic regression (McCullagh, 1980). The ordered logit model is commonly used in social science studies where features attitudes as the response variable. In general, the ordered logit model is most effective when the response is ordered with hierarchy and of discrete data types (e.g. attitudes: Content, Neutral, Upset).

$$\log \frac{Pr(Y \leq j)}{1 - Pr(Y \leq j)} = logit(Pr(Y \leq j)) = \alpha_{j0} - \eta_1 \times x_1 - ....... - \eta_p \times x_p \quad (2)$$

In some recent transportation studies, Naji et al. (2017, 2018) and Chang et al. (2016) have applied ordered logit models to study driving risks of near-crash events in China. In Naji et al. (2017), the authors first conducted a cluster analysis through a K-means algorithm to classify near-crash events based on their risk level and then carried out a traditional ordered logit model to find the contributing factors associated with the driving risk of near-crash events. Five impor-tant factors regarding environmental and driver attributes were found to be significant. A few months later, the same authors published another work (Naji et al., 2018) which used the same experimental setup for data collection and K-means clustering analysis as Naji et al. (2017). The only difference is that Naji et al. (2018) used a mixed-ordered logit model instead. The study was then able to find ten impactful factors associated with environmental and driver attributes, and five of the ten were new discoveries that the traditional ordered logit model based 2017 study missed out on. Chang et al. (2016) compared the ordered logit model to the mixed-ordered logit model on the same dataset and found 13 impactful factors with the traditional MNL model and 18 with the mixed model. The study also cited better performance metrics with the mixed model. The finding does not come as a surprise, since the mixed logit model is known for its ability to address unobserved heterogeneity issues and the potential variations in the effects of contributing factors. While the aforementioned studies adopted a similar data processing method to set up their discrete choice modelling as we did, the similarity of our works ends here. To the best of our knowledge, there have been no other studies that used DCM to predict vehicular GHG emissions at a road network level, making this study the first of its kind. Finally, to avoid any confusion, we should note a key point that the decision-maker in our DCM model is the link and not any human-centric decision-maker.

*3.2. Data Processing*

As the GHG emission values coming from the sensors and other monitoring systems are primarily continuous, we have to first discretize them into groups of categories. Ideally, an official governmental or inter-governmental guideline on determining the threshold for splitting high-medium-low levels of emissions would be the most meaningful method to base the discretization on. At the time of this writing, no such official guidelines exist. A viable alternative option is to use a data-driven approach and apply clustering analysis (Landau et al., 2011; Witten et al., 2016) to split the data into bins based on the underlying data structure. The common clustering methods are K-means and Hierarchical clustering, where K-means is more widely used than hierarchical clustering due to its ability to guarantee convergence into a predefined number of clusters. Since we already know we want a high-medium-low three-way split, K-means is a more suitable choice.

In this work, we are using the same dataset as Alfaseeh et al. (2020). Although we should note that our main purpose differs from Alfaseeh et al. (2020) who used the auxiliary-LSTM based continuous GHG emission prediction for eco-routing strategies, whereas we want to inform policy by providing the high-level decision-makers (policymakers) immediately interpretable and actionable information through categorical responses. All the variables used are explained in Table 1 and shown in Figures 1 and 2. All of the variables, except GHG Emission rate, are generated by the same microsimulator as in Alfaseeh et al. (2020). The time duration associated with the data is 1 hour long. The traffic conditions are assumed to be the condition of the start of an average working day's morning rush hour in the Downtown network of Toronto. The raw dataset features 72,960 data points. Within the dataset, there are a total of 12 variables containing information regarding the speed, density, flow, and road geometry. The most important variable in this dataset is vehicular GHG emission rate in links. Note that the GHG emission rate variable included here is an aggregation of several different types of Greenhouse Gases' emission rate. A subset of 5,000 data points was taken from the raw dataset to be used as a training dataset for the discrete choice model. Another distinct subset of 1,000 data points was used as test data to evaluate and compare the discrete choice model's prediction to the LSTM's results.

Table 1: Description of Model Variables

| Parameter | Description |
|---|---|
| Scenario | Traffic Demand level [The higher the number the more vehicles there are on road] |
| Link Number | ID associated with each link (road segment) in the road network |
| Time | Measured as time steps from beginning of the simulation to the end |
| Free Flow Speed | Average speed that vehicles would travel without congestion or other adverse conditions |
| Number of Lanes | Number of Lanes at the particular Link |
| Link Speed | Average Speed that vehicles travel at the particular Link |
| Link Toal Density | Total Density of vehicles at the particular Link |
| Link Density per Lane | Average Density of vehicles at each Lane of the particular Link |
| Link Total Flow | Total Flow of vehicles at the particular Link |
| Link Flow per Lane | Average Flow at each Lane of the particular Link |
| Delay on Link | Average time delay at the particular Link |
| In-Links Density per Lane | Average Density of vehicles at each Lane coming into the particular Link |
| In-Links Total Flow | Total Flow of vehicles coming into the particular Link |
| In-Links Flow per Lane | Average Flow of Vehicles at each lane coming into the particular Link |
| Flow over Capacity | The ratio of Flow over Capacity at the particular Link |
| GHG ER g/sec | Aggregation of multiple Greenhouse Gases' emission rate at the particular link. Measured in grams per second |

Before we attempt any analysis on the data, we first need to make sure that all the variables are within reasonably the same range to avoid misinterpretation on important information in the analysis due to scale differences. In other words, the data needs to be normalized first. In this study, we will be carrying out the normalization using the MinMaxScalar (MMS) method from the popular python machine-learning library sci-kit-learn. The MMS normalized the data by transforming the parameter values into the range of 0 to 1. However, many other transformers in the sci-kit-learn library also normalize the data to the 0 to 1 range. Compared to other transformers, MMS can keep the distributions of each parameter's data points as they were before the normalization (Figure 1), which is useful since the parameter data distributions can contain important information for the actual analysis results. Now that we have normalized the data into the same range, we can start understanding the underlying correlations amongst the parameters. A heatmap (Figure 2) was generated to visualize the correlations that each parameter has with the core target "GHG ER g/sec", which stands for GHG emission rate in grams per second. Looking across the heatmap, we note that the Link Speed (km/hr), Free flow speed (km/hr), and Link Density per Lane (veh/km) are the three most correlated variables to GHG emission rate. Such findings agree with the common consensus in the transportation field that vehicular emission level is most sensitive to vehicular speed (Ministry of Transportation, British Columbia. 2007 Delcan (2007)).

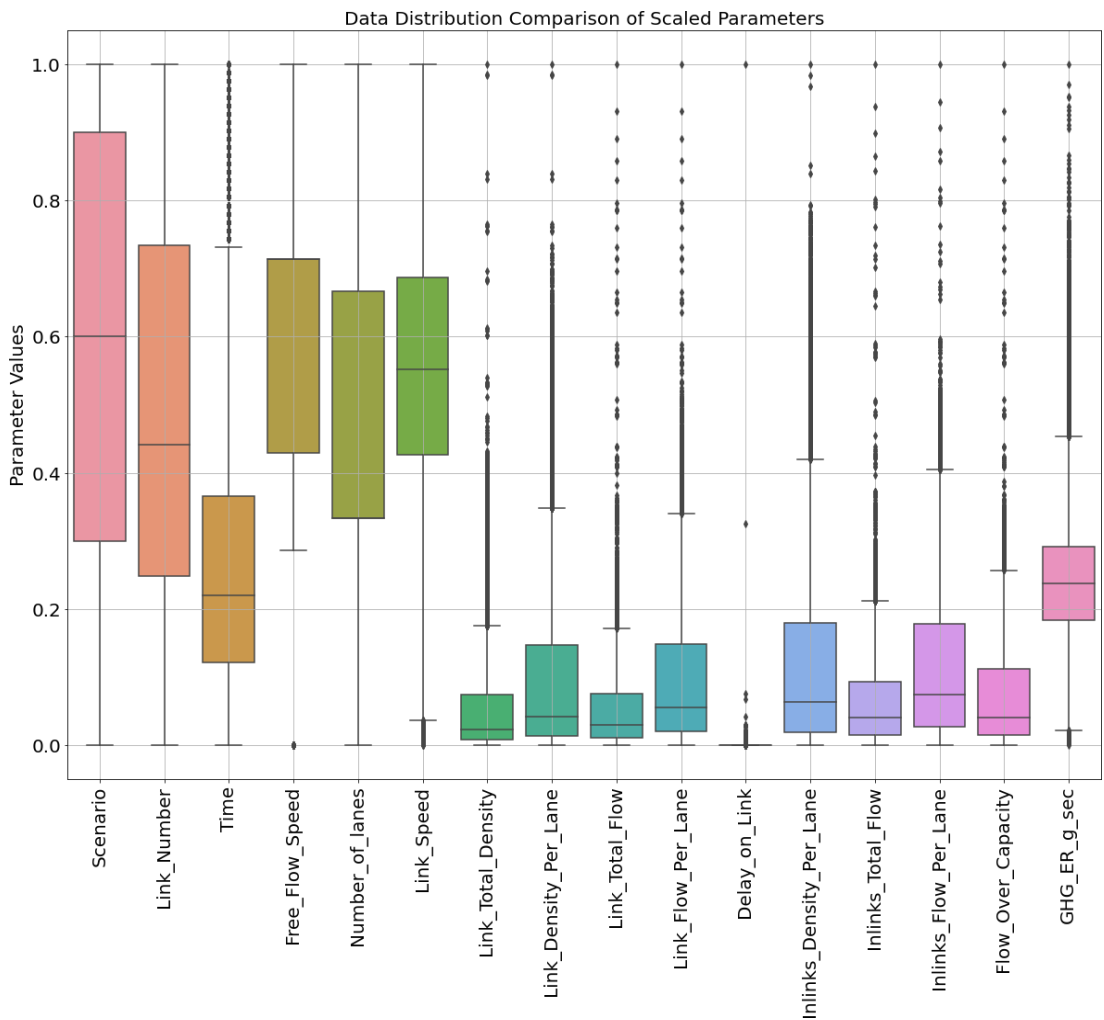

Figure 1: Using MinMaxScalar Transformer to normalize the data while preserving each parameter's distribution.

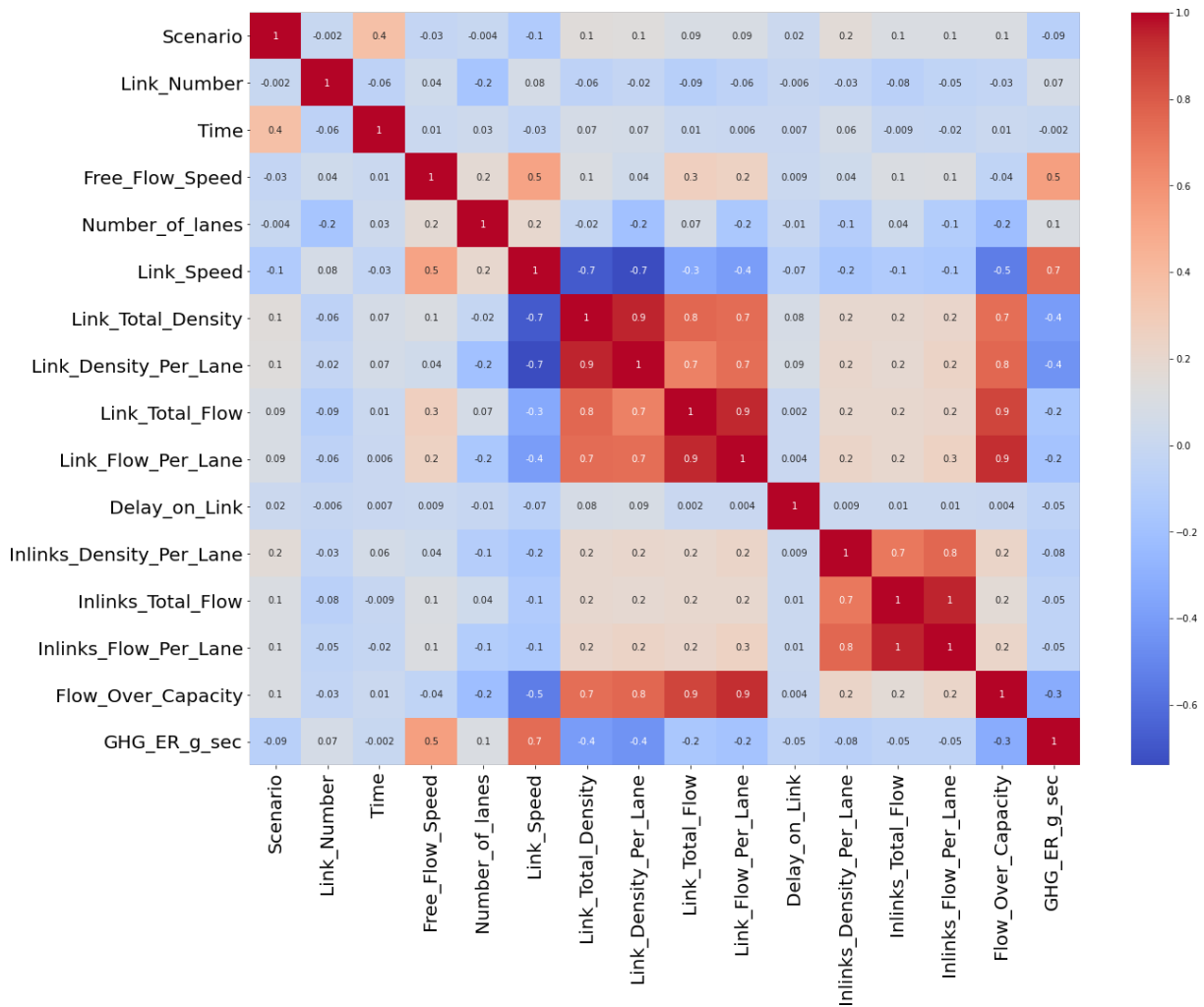

Figure 2: Correlations amongst the variables

However, correlation does not always convey the complete picture. There are more dimensions to the link-level GHG emission than the classic traffic variables such as speed, density, and flow that are included in this dataset. Link geometry, for one, also contributes significantly to the GHG concentration in links. Due to the limitation of the simulation, the only link geometry variable we have available here is "number of lanes". So, despite it not having as large of a correlation to the GHG emission rate in this dataset as some other variables, we are still going to include the "number of lanes" in our model to add the link geometry dimension. We are aware that there are other significant variables about link geometry that could have a significant impact on emissions such as link length and elevation/topology. Additionally, road type could also have an impact on GHG emissions. Another dimension we would like to bring in is the physical built environment. A common consensus found across relevant studies suggests that a near-link environment casts a large influence on the in-link emissions. Unfortunately, none of these variables are currently available in this dataset. If we could add these additional variables' information into our simulation, we would then be able to add them into our future analysis and potentially improve the prediction result even further. Since the utility function is linearly additive, the more variables we bring in, the better the prediction results. Nevertheless, to illustrate the idea of the general analysis framework and the effectiveness of our model, the variables that we are using are sufficient for now.

Apart from scaling the variables and identifying the proper model parameters, discretizing the GHG emission rate data into interpretable and actionable categories (i.e., low, medium, and high) is another crucial step before we conduct our discrete choice analysis. Normally, to give more real-world meaning to the choices, we would need to consult the government regulations on the cut-off thresholds of different levels of emissions. However, as of writing, to the best of our knowledge, there is no governmental documentation on within what range of values would GHG emission rate be considered as acceptable. To compensate for the lack of official governmental indicators, in this work we will be using a data-driven technique to come up with a high-medium-level of GHG emission rate.

Clustering analysis is an automatic process that divides data into similar groups with meaningful implications. Together, those groups provide multiple profiles of attributes. The profiles, in turn, allow us to find underlying patterns of input data. In this work, we lack the background knowledge of what is a good categorization of low-to-high emission levels. The data-driven clustering then becomes a useful technique to us since it can automatically detect similarities and correlations amongst the parameters and come up with a more comprehensive split than we could with a limited background. In general, there are two popular methods to go about clustering the data: K-means and Agglomerative clustering. Since we have a clear idea of the number of bins in mind, it makes more sense to feed the parameter into K-means to have it generate the exact number of bins (3) that corresponds to low, medium, and high levels of emission for us to interpret. As shown in Figure 3, the K-means method was able to come up with a clear-cut discretization for the GHG emission level. Each GHG emission rate data point was then assigned a number (1, 2 or 3) corresponding to the low-medium-high emission category that it falls into. At this point, the discretization is done, and we can start building the discrete choice model.

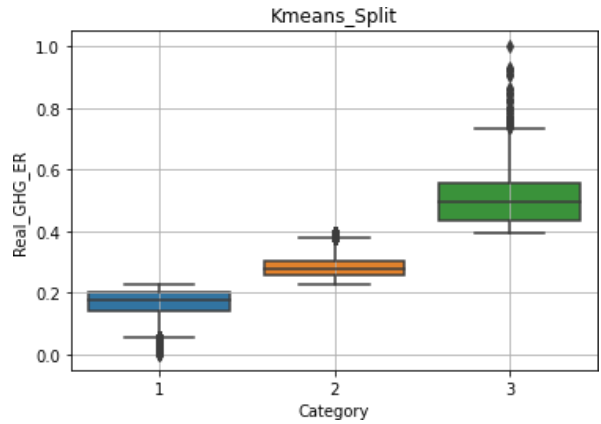

Figure 3: K-means based discretization: low-medium-high (1-2-3) GHG emission level.

### 3.3. Systematic Utility Specification

When we are working with Multinomial Logit Models, having robust utility functions is essential to guarantee the model's performance. In this work, we are working with categorical alternatives, hence it is only appropriate to have alternative specific parameters that reflect each category in each of the utility functions. Our main base model in this work is a dynamic MNL model. It is built with the most representative traffic variables and link geometry variables. We introduced the dynamic effect into the model by associating GHG emission level at each time step "t" with the emission level from the previous time step "t-1" in a Markov-Chain-like manner (A basic dynamic MNL utility function is shown in Equation 3 to illustrate the process). A total of three utility functions (Equations 4,5 and 6) were created for the dynamic MNL model to represent high-medium-low levels of GHG emission in links as alternatives. A single utility function for dynamic Ordered Logit model was also created (Equation 8).

$$V^*_{i,t} = ASC_i + \beta_x * X + \beta_{i,v,t} * V_{i,t-1} \qquad (3)$$

Regarding the setup of parameters and coefficients in the three utility functions (one for each emission level), we first highlight that the low-emission alternative was selected as the base case to investigate the scenario: if the low emission level was not chosen, what is the likelihood that medium/high level of emission would be chosen? So, for V1 (low emission) we are only estimating its Alternative Specific Constant (ASC) and not bringing in any Beta coefficients or other traffic parameters. To compensate for the setup of V1, the ASC for high emission was chosen to be set to zero. The reason behind such constraints on the utility functions arises from the limitation that for the K set of choices, the MNL model will only return K-1 sets of ASC and K-1 sets of attribute coefficients. With this setup (Equation 4,5 and 6), the MNL model would be able to return all of the alternative specific $\beta$ parameters as well as $ASC_{medium}$, and $ASC_{high}$ (refer to Table 2). And despite Equation 4 only consisting of a single constant, the model still works out for our purposes because we are fully exploiting the special characteristic that alternative specific constant (ASC) contains information regarding all the unobserved factors in the utility function by taking an average value of the unobserved factors. As a result, we would still have all the necessary information we need going into the predictions.

$$V_1 = ASC_{LOW} \tag{4}$$

$$\begin{aligned} V_2 = &\; ASC_{MEDIUM} + \beta_{MEDIUM\ SPEED} * Link\ Speed \\ &+ \beta_{MEDIUM\ DENSITY} * Link\ Density\ per\ Lane \\ &+ \beta_{MEDIUM\ FREESPEED} * Free\ Flow\ Speed + \beta_{MEDIUM\ NUM\ LANES} * Number\ of\ lanes \\ &+ \beta_{MEDIUM\ GHG} * Medium\ GHG + \beta_{HIGH\ GHG} * High\ GHG \end{aligned}$$

$$\tag{5}$$

$$V_3 = ASC_{HIGH} + \beta_{HIGH\ SPEED} * Link\ Speed + \beta_{HIGH\ DENSITY} * Link\ Density\ per\ Lane$$
$$+ \beta_{HIGH\ FREESPEED} * Free\ Flow\ Speed + \beta_{HIGH\ NUM\ LANES} * Number\ of\ lanes$$
$$+ \beta_{HIGH\ GHG} * HIGH\_GHG + \beta_{MEDIUM\_GHG} * MEDIUM\_GHG$$

(6)

Another attention-worthy part of the utility function defined above is the two new dynamic categorical variables that we introduced into the utility functions beyond the other readily available parameters on speed, density, and link geometry that were chosen directly from the dataset (shown in Fig 1 & 2). The two new variables, "Prev Med GHG" and "Prev High GHG", bring in time dependency and make GHG emission levels from prior steps parts of the exogenous variables. In a Markov process fashion, we first take the available GHG emission levels that were previously discretized into high-medium-low categories and offset them by one time step so that the GHG emission level at each time step t is associated with its previous time step t-1. We then separated the prior GHG emission level into two binary-entries parameters. One of the parameters indicates whether the "t-1"-th time step was medium emission Level or not (1 or 0), and the other parameter determines the same binary scenario for High emission levels. Note that there is no such parameter for low emission level because we have set that to be the base case for comparison.

The utility functions were then used on the test set to predict the choice probability/likelihood of the corresponding emission level with Equation 7, where "i" corresponds to a specific level (1/2/3, low/medium/high) and J = 3.

$$Pr(i) = \frac{exp(V_i)}{\sum_{j=1}^{J} exp(V_j)} \quad (7)$$

Once we obtained the probabilities of low, medium, and high emission levels for each data point, we take the category with the maximum probability and assign the data point the corresponding categorical number. Recall we established earlier that the decision-maker is the link. So, finding the most likely alternative means given the information of a data point, what is the most probable emission category that the link will choose as the alternative. After all the data points in the test set have been assigned a number, we could then draw a direct comparison with the validation set and check for the model's predictive accuracy.

The setup for the Ordered Logit model's utility is more straightforward as the standard form only uses generic parameters, resulting in a single systematic utility function for all the choices. It is built linearly under one single equation:

$$U = \eta_{SPEED} * \text{Link Speed} + \eta_{DENSITY} * \text{Link Density per Lane}$$
$$+ \eta_{FREESPEED} * \text{Free Flow Speed} + \eta_{NUM\ LANES} * \text{Number of lanes} \quad (8)$$
$$+ \eta_{MEDIUMGHG} * \text{Medium GHG} + \eta_{HIGH\ GHG} * \text{High GHG} + \varepsilon$$

The endogenous variable has a natural hierarchy that was predefined manually into 3 levels (low-medium-high). In the outputs, the model estimates all the coefficients $\eta$ (similar to $\beta$ in MNL model), as well as giving us two threshold values $\mu_1$ and $\mu_2$ that separates low/medium and medium/high. With these estimated coefficients and threshold values we can then predict Pr(GHG emission<=j). From here, we can get the probability for high-medium-low emission levels of each entry and check for the predictions' accuracy similar to the MNL model.

Finally, after the probabilities for the choice alternatives have been made and all the coefficients have been determined, we can use those information to conduct elasticity analysis [7] per:

$$E_{P_n(i), X_{ink}} = [1 - P_n(i)] X_{ink} \beta_k \quad (9)$$

Here, $P_n(i)$ is the choice probability for a certain alternative, $X_{ink}$ is a certain attribute (parameter), and $\beta_k$ is the coefficients estimated from the utility functions. In general, the elasticity analysis conveys key information regarding the most significant contributor(s) to a certain alternative being chosen as the choice alternative.

---

[7]Elasticity is a concept originated from Economics. It measures the percentage change of the target variable with respect to the percentage change of another variable.

## 4. Results and Discussion

In this section, we present all the significant results from the DCM models, including parameter estimations, general statistical metrics, prediction accuracy, and direct elasticity. We also discuss the implications of the findings presented. Finally, we conclude the section by proposing a couple of viable solutions to further improve the models.

After fitting the parameters into the utility functions and estimating them through the logit models, we obtained the parametric results outlined in Table 2 for both the MNL model and the Ordered Logit model. We observe that most of the parameters have a rather low Rob. P-value, indicating they are all significant variables to the model. The exceptions are the three variables based on number of lanes, and Link Density per Lane. As previously explained in section 3.2, number of lanes are the only link geometry variable we have hence it was incorporated in the model and kept despite the relatively larger p-value ($>0.05$). Similarly, since Link Density per Lane is a measure associated with number of lanes it was also kept in the model regardless of the comparatively higher p-value. Once we confirmed the parametric results, we then plugged the estimated parameters back into the utility functions to obtain the utilities. Using the utilities, we were then able to estimate the most likely outcome category/alternative with Equation 7. We should note that the Independence of Irrelevant Alternatives (I.I.A.) assumption amongst the choice alternatives (high-medium-low emission levels) were specifically tested for both the MNL model and the OL model using a mixed kernel test. The I.I.A. assumption holds for the choice alternatives in both models.

The general statistical metrics in Table 2 serve as a measurement for model comparison between the dynamic MNL and the dynamic OL. Since the model structure and the dependent variables used are not identical between the dynamic MNL and dynamic OL models, the Log-Likelihood ratio is the only statistical metric that could be appropriate for comparison. The LL ratio ranges from 0 to 1 and it reflects the efficiency of the model in terms of taking the same information and fitting a better model. The closer the LL ratio is to 1 the more efficient is the model fitting performance. In this scenario, the Dynamic OL model has a significantly higher LL ratio when compared to the Dynamic MNL, which suggest that the OL model is able to use more information in the data for their fitting. This finding confirms the general consensus that OL model is the better at handling Ordinal data. Since the exogeneous variable, GHG emission level, has an ordered hierarchy associated with it, the dynamic OL model is able to pickup that ranked relation better than MNL model and obtain a more efficient fit. Nevertheless, statistical metrics only serve as useful indicators, to further determine whether a model is performing better than another is very much still performance-dependent. Since one of the two key purposes of applying these models were to predict in-link vehicular GHG emission level, the accuracy of the predictions naturally should take more importance in the performance evaluation.

Table 2: Estimated Model Parameters for Dynamic MNL and Dynamic OL

| Parameter | Dynamic MNL | | | | Dynamic OL | | | |
|---|---|---|---|---|---|---|---|---|
| | Value | Rb std err | Rb p-val | Rb t-test | Value | Rb std error | Rb p-val | Rb t-test |
| $ASC_{Low}$ | 22.0 | 0.67 | 0.00 | 26.5 | | | | |
| $ASC_{Medium}$ | 11.8 | 0.59 | 0.00 | 17.1 | | | | |
| $\beta_{High\_Density}$ | -9.86 | 2.12 | 3.33e-06 | -3.42 | | | | |
| $\beta_{High\_FreeSpeed}$ | 18.6 | 1.14 | 0.00 | 12.2 | | | | |
| $\beta_{High\_GHG}$ | 0.74 | 0.22 | 7.82e-04 | 2.82 | | | | |
| $\beta_{High\_NumLanes}$ | 0.58 | 0.39 | 1.32e-02 | 1.45 | | | | |
| $\beta_{High\_LinkSpeed}$ | 11.7 | 1.24 | 0.00 | 6.68 | | | | |
| $\beta_{Medium\_Density}$ | 2.59 | 0.96 | 6.82e-03 | 2.57 | | | | |
| $\beta_{Medium\_FreeSpeed}$ | 3.11 | 0.56 | 3.29e-08 | 5.23 | | | | |
| $\beta_{Medium\_GHG}$ | 0.46 | 0.10 | 3.87e-06 | 4.52 | | | | |
| $\beta_{Medium\_NumLanes}$ | 0.48 | 0.20 | 1.68e-02 | 2.29 | | | | |
| $\beta_{Medium\_LinkSpeed}$ | 13.7 | 0.77 | 0.00 | 16.9 | | | | |
| $\mu_{low/medium}$ | | | | | 10.1 | 0.30 | 0.00 | 33.6 |
| $\mu_{medium/high}$ | | | | | 1.5 | 0.02 | 0.00 | 71.8 |
| $\eta_{Link\_Speed}$ | | | | | 11.1 | 0.656 | 0.00 | 16.8 |
| $\eta_{Link\_Density\_Per\_Lane}$ | | | | | -0.28 | 0.88 | 6.12e-0.2 | -0.315 |
| $\eta_{Medium\_GHG}$ | | | | | 0.18 | 0.09 | 0.00 | 1.96 |
| $\eta_{High\_GHG}$ | | | | | 1.02 | 0.16 | 0.00 | 6.25 |
| $\eta_{Number\_of\_Lanes}$ | | | | | 0.33 | 0.18 | 5.27e-0.2 | 1.87 |
| $\eta_{Free\_Flow\_Speed}$ | | | | | 5.78 | 0.51 | 0.00 | 11.3 |
| Log likelihood | -2498.51 | | | | -2527.34 | | | |
| LL Ratio | 0.0000142 | | | | 0.542 | | | |

With both the predicted results from the MNL model and OL model known, we generated two confusion matrices (Figure 4) with respect to the validation/test data. Addtionally, for the purpose of directly comparing the performance of the MNL and OL models to that of the LSTM model by Alfaseeh et al. (2020), we also converted the continuous regressional prediction results from Alfaseeh et al. (2020) into categorical responses through the same K-mean clustering method described in section 3.2. From here, we then generated a confusion matrix based on the LSTM results, shown in Figure 5. Looking across the confusion matrix for the MNL model (Figure 4 left), we see that (out of the 5000 sample data points) the MNL model was able to reach a consensus with the real emission categories up to 81.6% of the samples. The largest discrepancies between the predicted and the real GHG emission levels appear to be happening with the prediction of category 2 (Medium level of emission), where the MNL model was only predicting up to 79.4% of accuracy. Most of the disagreement occurred when the MNL model inaccurately categorized 17% mid-level GHG emissions into the high-level category, and followed by another 13% of mid-level emissions being classified into the low-level category. On the other end, the OL model's prediction (Figure 4 right) exhibits a better result than that of MNL. The OL model's prediction achieved up to 85.3% consensus with the real data, which is better than the 81.6% accuracy of the MNL model. The better performance of OL model is to be expected since the model specializes in handling ranked ordinal data as its niche. And in this case, the low-medium-high emission levels are ranked ordinal data. However, despite the anticipated better overall prediciton performance, the OL model also underestimated a great deal of the category 2 (mid-level) cases and mistook them for category 1 (low-level). Although the prediction accuracy for category 3 (high-level) cases of the OL model notably was able to achieve up to 92.9% accuracy. While the most important aspect of this prediction framework is to address high-levels of GHG emissions, the mid-level GHG emission's prediction performance is still another important topic that deserve attentions. In the prior study of Alfaseeh et al. (2020), the authors mentioned an issue in the prediction of GHG as a continuous variable, where their LSTM regression model was having trouble maintaining high accuracy when predicting for higher magnitudes of GHG emission rate. From the categorized LSTM model results shown in Figure 5 we also observe a similar issue of worse prediction accuracy at the medium-level. There could be numerous reasons behind the comparatively lower prediction accuracy that occurs to all three models when predicting for medium emission level of emission. In general, the problem is most likely contributed by the discretization process, which potentially introduced uneven variability across the three emission levels. As a result, high standard deviation was introduced to the clustered medium level. We can observe that all three prediction models (4 and 5) consistently underestimated a significant bunch of medium level data points to be classified under lower level despite the differences amongst the models. Thus, the lower

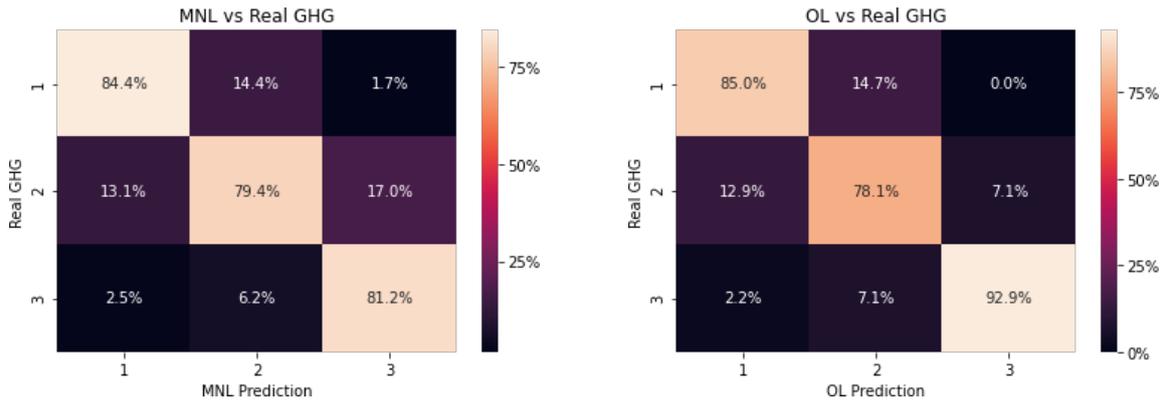

Figure 4: Side-by-side comparison of MNL model's and OL model's Confusion Matrices with respect to their emission level predictions vs Real GHG emission levels. Levels 1-2-3 (on the axes) corresponds to low-medium-high

accuracy (for medium level) is more likely an inherent issue from data processing (clustering) rather than a problem of the DCM or LSTM model. While the K-means clustering split is the best alternative data-driven solution we have to fill in the blank of the expert-opinion-based guideline, the thresholds and range of values defined by the discretization may still not be the most ideal. Despite the seemingly well-defined split that the K-means algorithm produced in Figure 3, we can see that values considered at the upper 50% of the low emission category and values considered at the lower 50% of the medium emission category have a difference of less than 0.1. In addition, we also see that there are outliers in category 2 crossing over into category 3. Ideally, an official guideline would help provide a more clear-cut split, which could help with the discretization. From a modelling perspective of DCMs, there is another limitation that might be hindering the overall emission prediction, which is the lack of certain informative variables in the model. There are likely variables closely associated with different levels of vehicular GHG emission not being featured in the model, as the ones mentioned in section 3.2 including the type of roads, additional road geometry variables like link length and topology, and built environment. These variables could be closely associated with certain emission level, and bringing them into the DCMs could help with the prediction for all 3 levels. Therefore, it is our goal to bring in as much of that additional information as possible in our future models.

Beyond the known additional information that we could bring in, to come up with further new variables one could look to see which variables in the current model influenced the choice of high emission level the most and try to extend along that line of information. A direct elasticity analysis would reveal just how much could each of the variables/attributes influence the choice of a certain alternative. We have generated two direct elasticity box plots, one for each DCM model, where high GHG emission levels were selected as the alternative (Figure 6). Based on the boxplots, the MNL model is suggesting that, of all selected variables, Free flow speed is the most impactful attribute associated with high levels of GHG emission being chosen, followed by link speed. Similarly, the OL model's direct elasticity identified link speed to be the most significant contributor followed by free-flow speed. Typically, with the increase of speed, the emissions will start to decrease to some extent, however, after internal combustion engine has been burning fuel at a higher rate to maintain high speeds for some time, the emissions tends to go back up. This trend was observed in a prior study by Djavadian et al. (2020a), hence explaining the positive values of the two speed variables in the direct elasticity. Nevertheless, the elasticity analysis findings that the only two speed variables available from the model are the top two contributor to high GHG emission levels point to the direction that having more relevant additional speed-related variables would likely help with the predictive modelling. We should note that this type of information is not only useful for us as modellers to help select more potentially meaningful variables to add on to improve the model prediction, which is only half of the motivation of using DCA here, but those are also extremely interpretable information to the policymakers. Since the MNL model and OL model are both linearly dependent to its predictor parameters, the larger the magnitude of parameter coefficients the more impactful the parameter. In terms of impacting the prediction for the most likely choice alternative (in this case: emission level), let us take the MNL as an example to explain the information encoded in the parameters. In the case of MNL model's maximum likelihood prediction for GHG emission level at high level, we know that Free Flow Speed shows the strongest positive correlation. That means, if Free Flow Speed is high, chances are the GHG emission level is going to end up in the high emission level category. So should those resourceful agencies decide to take immediate actions and curb down the high GHG emission levels, they can look at this finding from the elasticity analysis and easily identify that the most significant factor is Free flow speed. With this information of key contributors to high emissions, the decision-makers can then create specific policies tailored to regulate free-flow speed on roads and directly cut down the chance of having high GHG emission levels on road. The most common example of such policies is the traffic-calming-measures, where they set appropriate speed limits or different speed zones with respect to the emission levels, while taking into account of the differences in the road functional classes (Minor, Collector, Local, etc) that corresponds to each of the road segments. Such policies will keep the speed in a sweet spot somewhere between being too low that leads to congestion, or being too high that induces more engine exhausts.

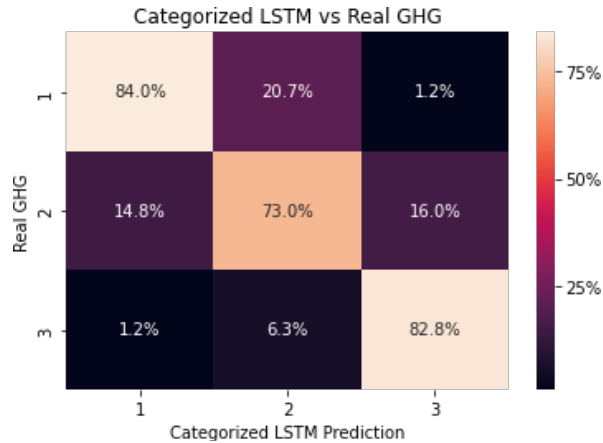

Figure 5: Confusion matrix generated from LSTM regression model's results of Alfaseeh et al. (2020) for performance comparison with DCM models

The ease of interpretation that comes with these DCM model-based elasticity analyses can be translated into actions directly. The smooth transition is built on the fact that the DCM models can easily convey information to the user, which is a key aspect that powerful state-of-the-art prediction tools such as LSTM lack. Another technical observation we made, that is quite useful for interpretation, regarding both elasticity analysis, is that the attributes with the largest amount of variance are the most influential ones to the predicted alternative. When we cross compared the parameter to model correlation traits to the elasticity analysis, we can see (Figure 5) that those more impactful parameters also exhibit greater variance in their data value distributions compared to the other parameters. Even without reading the information of Table 2, we could still obtain the information about the impactful attributes in the model based on examining the direct elasticity comparison charts.

With respect to the model performance, by comparing Figure 4 and 5 we can see that the DCM models managed to achieve an overall higher prediction accuracy (around 81% for MNL and 85% for OL) than the Auxiliary-LSTM model (around 78%) used by Alfaseeh et al. (2020), where the authors also cited a 77% model accuracy in their paper, using the same data set as we used in this work. Based on pure predictive power, in general, the LSTM no doubt has the better accuracy. Nonetheless, in this case of working with discrete data, our DCMs managed to edge out the results extracted from the LSTM models that were originally used for predicting continuous data. It is important to point out that the findings of our result is not to say one model is more superior than another. With respect to different purposes of use, one model might be more suitable than another. Alfaseeh et al. (2020)'s goal was to get better Eco-routing strategies through a multi-objective optimization process, and working with continuous data was more convenient. Whereas, we want to inform policy and actions by providing high-level decision-makers with an immediately interpretable and actionable analysis. Hence, for our purpose, working with discrete data is more suitable.

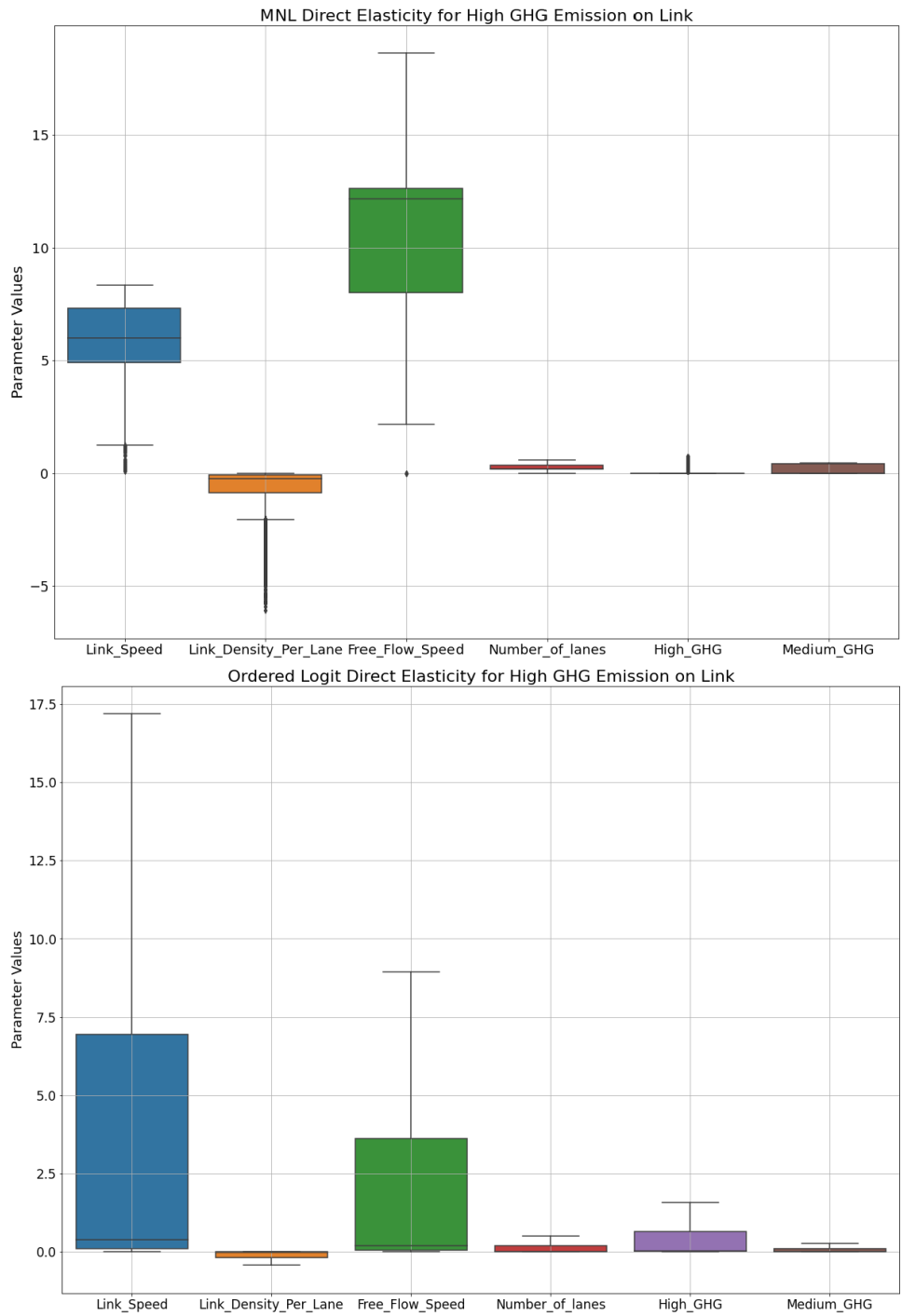

Figure 6: Comparison of MNL model's and OL model's Direct Elasticity with respect to High GHG emission level predictions.

Nevertheless, as an additional future application scenario of our work, we could experiment with integrating our proposed GHG prediction framework into the objective functions of other road network-level multi-objective optimization problems that needs to account for GHG emission-reduction to see if our framework can result in a more sustainable low-GHG emission solution. As a potential candidate, Meshkani and Farooq (2022) recently introduced a novel ride-matching algorithm for on-demand shared mobility where one vehicle can serve multiple passengers. Their proposed algorithm can provide cities with a sustainable and clean transportation mode as it can decrease the vehicle kilometer travelled and vehicle congestions, which would then lower emissions. In our work, we are using the same base simulation as the one used in Meshkani and Farooq (2022). There, the objective function for optimized matching was mainly based on Vehicle Kilometer Travelled and Congestion. However, as an additional future application scenario of our work, the prediction framework proposed here could be integrated into their objective function (and its optimization) to introduce another dimension of eco-matching in the algorithm. Thus, the matching algorithms can then directly account for the emission with our model, which could help further improve Meshkani and Farooq (2022)'s algorithm to be even cleaner and more sustainable. However, it should be noted that this extension is not in the current scope of our work, but a potential future application.

## 5. Conclusion

In this work, we examined the possibility of using DCM as an alternative way to directly predict GHG emission level for vehicles in links, with the emphasis on making use of the model's interpretability and actionability to provide insights to high-level decision-makers. We tested two dynamic DCM models, namely dynamic MNL and dynamic OL. The models were then used to make predictions on potential emission levels and the prediction accuracy was measured against the simulated real vehicular GHG emission data. The prediction accuracy of the GHG emission rate of the DCM models was compared to that of a state-of-the-art Auxiliary-LSTM model Alfaseeh et al. (2020), which used the same data set as this work. Despite the Auxiliary-LSTM being the more complex and robust prediction model, in this study, the dynamic DCM models managed to outperform the Auxiliary-LSTM for discrete response variable (GHG emission) in prediction accuracy by more than 5% (MNL 81.6% & OL 85.4% to LSTM 78.6%).

Another important contribution of this work is the framework of discretizing continuous data to make a categorical prediction to produce interpretable analysis that induces direct actions from decision-makers. The interpretability comes from the simplification of reducing the continuous GHG emission rate into 3 discrete levels of categories, each of which corresponds to a level of actions needed from the decision-maker, with the highest level of emissions demanding immediate actions to be put in place. Meanwhile, for medium-level of emission, some actions can be taken by the decision-maker but not in the most urgent manner. The lowest level of emission calls for no necessary actions from the decision-maker. By these levels of urgency and corresponding strategies, the framework also brings actionability. In the environment of an urban intelligent transportation system, those live link-level emission information provides a dimension of prioritization for managing the entire road network constituted by all the links. The decision-maker can choose to first address links with high emission levels by, for instance, changing the adaptive control traffic signals in the network to redirect flow from high emission level links to low emission level links and maintain a free flow speed that is within the acceptable emission range (low to medium). The decision-makers will then only deal with links with a medium level of emission if no high level of emission exists in the network. This is similar to the practice commonly used in road infrastructure asset management where only the level of road conditions (Good, Mediocre, Poor) in a network needs to be known to asset managers to help decide what immediate actions (No repair, Minor repair, Major repair) need to be taken and for which road.

Furthermore, elasticity analysis for both DCM models was carried out after the predictions were made to examine the influence that each attribute has on causing high GHG emission levels. The analysis shown that for both MNL and OL model, Free Flow Speed and Link Speed are the top 2 most impactful factors for high GHG emission levels. The positive parameter values for those two factors indicates that as the link speed and free flow speed increases, the chances of having high emission level on the road link also goes up. The finding also confirms the known physical phenomenon that as the speed goes up, the combustion engine in the vehicle would have to work harder to sustain the high speed and creating more exhaust gas as a result. The information from the elasticity analysis can enable policymakers/urban planners (or other resourceful solution seekers) to easily assume the role of decision-maker and tailor solutions (e.g. traffic calming zones) to directly curb the most significant contributors to high emission levels and prevent them from contributing to more emissions from the get-go.

Finally, we recognize that since the utility function in DCM is of linear nature, its performance depends heavily on having as many impactful independent variables as possible, we will bring in variables from dimensions such as the physical built environment of the links and near the links to make up for the important missing information. So the model would not have to rely heavily on alternative specific constants for each alternative's utility function. Other models such as mixed logit models can also be used in conjunction with the dynamic MNL and OL models to address unobserved heterogeneity issues and potential variations in the effects of contributing factors. A mixed model could also help us conduct meaningful cross-elasticity analysis for the MNL model, which is otherwise impossible if the MNL model is to be used on its own. Nonetheless, overall, the dynamic DCM models in this work have proven to be very informative, easily interpretable, and highly actionable to decision-makers. The prediction accuracy of the dynamic DCM models also exceeded our expectations and proved to be very capable. While there is still room to improve, we believe that, with the findings in this work,

DCA-based framework can be adopted as a convenient tool for GHG emission predictions and reference of corresponding mitigation strategies for decision-makers in the urban transportation settings.

**Declaration of Competing Interest**

The authors declare that they do not have any conflict of interests associated with the submission.

2021.

H.-T. Pao, Y.-Y. Li, and H.-C. Fu. Causality relationship between energy consumption and economic growth in brazil. *Smart Grid and Renewable Energy*, 2014, 2014.

D. Radojevic´, V. Pocajt, I. Popovic´, A. Peric´-Grujic´, and M. Ristic´. Forecasting of greenhouse gas emissions in serbia using artificial neural networks. *Energy Sources, Part A: Recovery, Utilization, and Environmental Effects*, 35(8):733–740, 2013.

J. E. Rito, N. S. Lopez, and J. B. M. Biona. Modeling traffic flow, energy use, and emissions using google maps and google street view: The case of EDSA, philippines. *Sustainability*, 13(12):6682, jun 2021. doi: 10.3390/su13126682. URL https://doi.org/10.3390%2Fsu13126682.

D. Stanek and C. Breiland. Quick estimation method for greenhouse gas emissions at intersections. Technical report, 2013.

I. Witten, E. Frank, M. Hall, and C. Pal. *Data Mining: Practical Machine Learning Tools and Techniques*. The Morgan Kaufmann Series in Data Management Systems. Elsevier Science, 2016. ISBN 9780128043578. URL https://books.google.ca/books?id=1SylCgAAQBAJ.

Y. Yao, X. Zhao, C. Liu, J. Rong, Y. Zhang, Z. Dong, and Y. Su. Vehicle fuel consumption prediction method based on driving behavior data collected from smartphones. *Journal of Advanced Transportation*, 2020:1–11, 2020.

J. Zhao, J. Zhang, S. Jia, Q. Li, and Y. Zhu. A mapreduce framework for on-road mobile fossil fuel combustion co2 estimation. In *2011 19th International Conference on Geoinformatics*, pages 1–4, 2011. doi: 10.1109/GeoInformatics.2011.5980759.
34